# Thermo-mechanical analysis of tumor-tissue during cryosurgery: a numerical study

G. Sai Krishna[1], Anup Paul[2*]

Department of Mechanical Engineering, National Institute of Technology Arunachal Pradesh, Jote, Arunachal Pradesh-791113, India

E-mail addresses: [1] saikrishna889@outlook.com, [2] catchapu@gmail.com (*Corresponding author)

**Abstract**

Since decades, many techniques have been developed for curing cancer. Here, an effort has been made to cure tumor by inserting a cryoprobe into the target region which is connected to a cryogenic mechanism. This technique is presently used only for curing simple tumors and hence the motive here is to trigger this technique towards curing large sized internal tumors. Thermo-mechanical study was done for the tumor-tissue subjected to cryotherapy using COMSOL Multiphysics™ software. A 3-D model has been presented considering all the properties to be a function of temperature. Phase transformation of bio liquid to ice was also considered in the study. Also, parametric study has been done by varying the probe and tumor dimensions and also by varying the tumor location. Pain tolerance of the tissue during this process was also studied. Variety of thermal and mechanical distributions have been extracted from the study which gave an idea about the time of operation, pain experience and the effect of process on the surrounding healthy tissue. It was found that, heat transfer and thermal stress in the tissue increased with increase in probe diameter and decrease in tumor size and reached a steady state at around 50 mins. Cryosurgery becomes simpler as the tumor moves out of the tissue. It was also predicted that the patient might experience severe pain by the end of treatment. This study might help researchers and surgeons in improving the understanding of cryosurgery process along with its limitations.

## 1. Introduction

Cancer is one of the major death causing diseases in the world. According to WHO, around 10 million deaths occurred in the year 2020, that is nearly one sixth of total deaths [1]. ICMR reported [2] 13.9 lakh cancer cases in 2020 and estimates an increase to 15.7 lakh by 2025 in India. The number of cancer patient is increasing [3] at a rate of 0.4424% every year in the United States. It was also reported that one in every four citizens [4] of the United States die due to cancer. Tumor is a lump of unwanted cells which grow rapidly and spread over healthy tissues. The diseased cells start growing within any tissues, bones and glands. Cancerous tumors are the lump of cells which invade and dominate the growth of healthy tissues by consuming all the essential nutrients and become insensitive to the immune system of the body and affect the functioning of healthy tissues which may even lead to death. 2-D Ultrasound Imaging, X-ray, Positron Emission Tomography (PET) scan, Computed Tomography (CT) scan or Magnetic Resonance Imaging (MRI), etc. are the different techniques used to detect tumors.

Cryosurgery is a thermal ablation technique to kill the tumor cells using the science of cryogenics. Cryogenics is the study of generating the temperature below 120K (i.e. $-153^0C$) which is achieved by expanding high pressure liquified gas such as nitrogen, hydrogen, helium, etc. whose evaporating temperature is around $-190^0C$ (may be still lower). During the process of cryosurgery, this cryogenic liquid is made to pass through the cryoprobe and the cell temperature is severely reduced and maintained by inserting this cryoprobe into the target region due to which the liquid in the cell freezes by transferring the heat from the tissue to the cryogenic fluid. The complete denaturation of a cell occurs at a particular temperature depending on the cooling rate of the process. Here, temperature of $-40^0C$ [5–7] is chosen as the lethal temperature during the process. To achieve this temperature in the target region, the cryoprobe is made to attain a temperature around $-185^0C$ and made to have a good thermal contact with the target region [8] and therefore this process is named as Cryotherapy or Cryoablation or Cryosurgery. Causes of cell death can be (a) chemical- using drugs and toxins. (b) biological- due to viral infections or malfunctioning of cytokines. (c) physical- due to ionized radiation or thermal excitation or mechanical stress. (d) natural cell death.

In cryosurgery, destruction of the tumor cells is physical which occurs because of the formation of ice ball in the target region. This results in obstruction of blood circulation in the frozen region, known as necrosis and then generates mechanical forces in the target region and results in chopping off the cell organelles. At last, it results in apoptosis i.e., cell death which occurs due to the extreme irregular conditions in the surroundings. Generally, these cells enter the apoptotic state within 8 hrs [9] after rewarming after cryotherapy.

During the process there are four major phases [10,11] which are formed in the tissue while freezing: (a) Kill region (Due to the absence of blood flow and metabolic activity, cells in this region gets destroyed at a temperature below $-40^0C$), (b) Frozen region (All the liquid gets converted into solid hence forms an ice ball. The maximum temperature is around -$10^0$C), **(c)** Mushy region (It is the region where the tissue material is in transition state, temperature range is in between $T_{solidus}$ and $T_{liquidus}$. $T_{solidus}= 0.5°C$. $T_{liquidus}=-10^0C$), and **(d)** Unfrozen region (It is the region where the influence of cryoprobe is recognised but remains unfrozen).

The phenomenon of solidification of cellular liquid is known as cellular crystallization, this occurs in two different ways, extracellular crystallization and intracellular crystallization. Extracellular crystallization is also known as the solution effect which is one of the causes of cell destruction. This occurs when the tissue is cooled slowly in the range of $-4^0C$ to $-21^0C$ due to which water migrates from the cell interior to exterior and therefore ice is formed in the extracellular space. This causes mechanical damage to the cell membrane due to crushing of the cell boundary at low temperature. Intracellular ice formation is the primary cause of cell destruction. This occurs when the cooling rate is rapid in the temperature range of -15 to $-40^0C$. Since there is no time for the transfer of cellular liquid to extracellular space, crystallization of the cellular liquid occurs in the internal space itself, which is highly unstable [9] and results in the destruction of the cell.

Cryosurgery has few major advantages which makes it unique from other techniques [9], such as:

I. Unlike other surgery processes, there is very mild bleeding of body fluids as everything gets frozen. This is a great advantage because blocking the body fluid flow during surgery is a hectic task for surgeons.

II. After the tumor cells die, they automatically get divided and scars out from the healthy tissue with high cure rate.

III. The treatment is less painful compared to other treatments.

IV. Comparatively easy to perform with low cost.

Study was also done to analyse the pain in the tissue. Pain is an uneasy sensation in the body due to external attacks or internal malfunctioning of body organs. Pain-tolerance is a highly relative quantity which cannot be standardized in general as it changes from person to person and also changes with respect to the external and psychological conditions [12] through which the person has gone through. But a few rough analyses have been made through medical trials to know the average pain threshold limits that a human body can bear. It was found that different organs of the human body have different threshold limits of pain pressure. Pressure threshold is the minimum value of the induced pressure that generates unbearable pain in the tissue. Although, maximum tensile strength varies from tissue to tissue ranging from 0.5MPa for arterial wall to 200MPa for bone [13], it is observed that bearable threshold stress among different human organs is much lower than the tensile strength of the respective tissue, ranging from 0.3MPa to 1MPa [14,15]. Considering a factor of safety of 3, 0.1MPa was chosen as the maximum bearable stress for the tissue in this study.

Very limited research [16] has been done till date in this field and is used only for curing simple and limited cases such as skin, eye, oral cavity tumors. Observing the uncompromisable advantages of the process, it is highly recommendable to extend this technique for curing cancer of internal organs and also for large sized tumors.

Researchers have worked intensely for the development of cryosurgery. Biological heat transfer was initially studied by Pennes [17,18] where he had analysed the temperature distribution on a resting forearm and analysed the relation between arterial blood flow and the temperature in the forearm tissue and then developed the governing equation for the biological heat transfer. Lipkin and Hardy [19] performed experiments for determining the thermal properties of human tissue. Major discoveries of cryogenics happened in the late 19$^{th}$ century by developing expansion systems for liquefied gases (Oxygen, Nitrogen, Hydrogen, Helium, Argon etc.). These advancements in cryogenics led to the rise of cryosurgery [7]. Initially the study of cryogenics was used for cryopreservation, to store body organs, bacteria etc. Subsequently scientists used the field of cryogenics for treating simple tumors in the human body. In the early 1960s, due to inefficient devices, this was limited for freezing a few millimetres of thicknesses only and therefore this therapy was used only for treating superficial layers of tissues, often in the fields of gynaecology and dermatology [7,9,16]. As a result of advancements in the imaging techniques, it created a path for the development of cryosurgery and other invasive and non-invasive techniques. Much advancement occurred in the 1960s. Mazur [5] in 1970 analysed the response of biological cells when subjected to cryofreezing. Literature [16] reveals that surgeons used cryosurgery for treating brain tumors and made clinical trials and predicted the possibility for treating the tumors in internal organs such as liver, lung, kidney, etc. Few authors also determined [20] the thermal stress distribution in the tissues numerically using commercial softwares. Rabin and Steif [21] have developed a mathematical model for the stress distribution for the freezing process in the tumor and the surrounding tissue. Study [7] of cell destruction through cryosurgery mechanism and development of cryosurgical systems happened in the late 20$^{th}$ century. Rabin, Y. and Steif, P.S. [22] updated their model for studying the stress distribution subjected to cryosurgery with subsequent thawing. Few scientists [23–26] made experimental trails by performing in-vitro experimental trials on tissue phantoms. Few authors also made in-vivo experiments, Seifart et al. [8] have conducted few experiments on pigs in-vivo in different working conditions to study the temperature distribution in the tissue during the cryosurgery process. K.J. Chua [25] made an in-vivo experiment on a pig- liver, considering blood flow in the tissue. He also performed experiments for analysing the effect of input parameters of the cryogenic fluid such as flow rate, pressure etc. on the process of cryosurgery. Authors [11,27] made several innovations and advancements like using multiple probes and hybrid probes for treating large sized internal irregular tumors. Liu et al. [28] have developed a minimally invasive probe system which can perform both cryotherapy and hyperthermia consequently for treating tumors which are deep inside the tissues. Researchers [25,29–31] have also made trials by injecting nanoparticles into the tissue for increasing the efficiency of the cryosurgery and other thermal ablation process. Several 2-D and few 3-D numerical models were developed [32–35] for studying and increasing the efficiency of the cryosurgery process and also for treating tumors in different internal organs. Authors [10,36,37] also tried to make these models more realistic by considering blood vessels, considering the thermal and physical properties of the tissue and the tumor to be a function of time and temperature. Few authors also considered the phase transition [38] that occurs during the process and improved the understanding of cryosurgery.

Cryosurgery is still in its initial stage of development, there are no specific guidelines for the surgeons for performing this surgery. Presently this technique is used for curing simple and limited cases such as oral tumors, skin tumors etc. Even research on mechanical study of the process is in its very initial stages and pain experienced during the process is not yet studied, to the best of author's knowledge. Considering the advantages of this cryosurgery technique (mentioned earlier), it is highly recommendable to develop this technique to its fullest extent and make it practically usable for treating internal and large sized tumors.

## 2. Mathematical model

### 2.1 Physical model

A 3-D domain was chosen where the dimensions have been determined by considering the average size of the human internal body organs like liver, lung etc. For the ease of the study, the tumor was considered to be a cube of side 20 mm and the tissue was considered to be a cuboid of length 90 mm, width 60 mm and height 60 mm. Probe is considered to be a cylinder of height of 23 cm with a conical tip of height of 2 mm and diameter 6.9 mm. Probe is inserted at the position where its tip is 4 mm above the bottom surface of the tumor and the tumor is placed at a location where the tumor's centre is 15 mm above the tissue's centre. Figure 1 shows the physical model of the tumor tissue with inserted cryoprobe.

<Fig. 1>

Both the tissue and the tumor materials were considered to be homogeneous and isotropic. All the thermal and mechanical properties of both tissue and tumor are considered to be function of temperature [38] as shown in the table-1. Only biological conduction mode of heat transfer with phase change was considered in the study.

<Table-1>

Initially, temperature of the tissue and tumor are maintained at $37^0$C, which is the core temperature of the body. Domain is assumed to be linearly elastic and also considered to be fixed at the bottom surface and body weight is also considered in the study.

Probe surface, from tip to a height of 11.5 cm is maintained at a cryogenic temperature which is a function of time [8], which reaches to $-180^0C$ in the first 3 minutes and roughly decreases by $5^0C$ within the remaining time. All the remaining surfaces are considered to be insulated. Probe surface is also assumed to be fixed. It is to be noted that all the observations are done at different time steps and also at different locations in the domain to study the variation of thermal and mechanical parameters with respect to space and time respectively.

**2.2 Governing equations**

General heat conduction equation (1) was considered for the case of 3-D unsteady model. Heat transfer in the biological media is governed by Pennes bioheat equation (2). Pennes equation is the modified form of general heat conduction equation, where heat generation due to blood perfusion '$Q_p$' and metabolic heat generation '$Q_m$' are replaced with the heat generation term '$Q$' in the heat conduction equation.

<Fig. 2>

$$\nabla \cdot k \nabla T + Q = \rho c \frac{\partial T}{\partial t} \qquad (1)$$

$$\rho c \frac{\partial T}{\partial t} = \nabla \cdot k \nabla T + Q_p + Q_m \tag{2}$$

$$Q_p = \rho_b c_b w (T_b - T) \tag{3}$$

$k$ represents the thermal conductivity, $c$ represents the specific heat, $\rho$ represents the density of the biological media and $w$ represents the blood perfusion rate.

$$c = \theta_1 c_1 + \theta_2 c_2 + L_{1 \to 2} \frac{\partial m}{\partial t} \tag{4}$$

$$m = \frac{1}{2} \frac{\theta_1 - \theta_2}{\theta_1 + \theta_2} \tag{5}$$

$$k = \theta_1 k_1 + \theta_2 k_2 \tag{6}$$

Here, fig. 2 represents the schematic of the phase change process from liquid to solid as defined by Comsol Multiphysics [39]. Here, $c_1, c_2$ are the specific heats, $\theta_1, \theta_2$ are the mass fractions, $k_1, k_2$ are the thermal conductivities of liquid and solid phase of biological media and $c$, $k$ are the overall specific heat and overall thermal conductivity. $L_{1 \to 2}$ is the latent heat of freezing of the biological media and $m$ is the phase transition function of the material that is undergoing the phase change. Initially whole system is in liquid phase ($\theta_1 = 1, \theta_2 = 0$) and gets completely transformed to solid phase ($\theta_1 = 0, \theta_2 = 1$) by the end of the process. Blue line represents the mass reduction of the liquid phase and the red line shows the increase in the solid phase. Phase transition occurs exactly when both the lines start turning i.e., in between the dotted lines for a temperature span of $\Delta T_{1-2}$ from $T_1$ to $T_2$ and phase change temperature $T_{pc}$ is assumed to be the average of $T_1$ and $T_2$. While implementing the phase transition function, a smooth transition is assumed between the phases, within the interval of phase change temperature.

$$\sigma_{xx} = \frac{E}{1-v} \varepsilon_{xx} + \frac{vE}{(1+v)(1-2v)} (\varepsilon_{xx} + \varepsilon_{yy} + \varepsilon_{zz}) - \frac{\alpha E}{1-v} \Delta T \tag{7}$$

$$\sigma_{yy} = \frac{E}{1-v} \varepsilon_{yy} + \frac{vE}{(1+v)(1-2v)} (\varepsilon_{xx} + \varepsilon_{yy} + \varepsilon_{zz}) - \frac{\alpha E}{1-v} \Delta T \tag{8}$$

$$\sigma_{zz} = \frac{E}{1-v} \varepsilon_{zz} + \frac{vE}{(1+v)(1-2v)} (\varepsilon_{xx} + \varepsilon_{yy} + \varepsilon_{zz}) - \frac{\alpha E}{1-v} \Delta T \tag{9}$$

$$\sigma_a = \frac{\sigma_{xx}+\sigma_{yy}}{2} \pm \sqrt{\left(\frac{\sigma_{xx}-\sigma_{yy}}{2}\right)^2 + (\tau_{xy})^2} \qquad (10)$$

$$\sigma_b = \frac{\sigma_{yy}+\sigma_{zz}}{2} \pm \sqrt{\left(\frac{\sigma_{yy}-\sigma_{zz}}{2}\right)^2 + (\tau_{yz})^2} \qquad (11)$$

$$\sigma_c = \frac{\sigma_{zz}+\sigma_{xx}}{2} \pm \sqrt{\left(\frac{\sigma_{zz}-\sigma_{xx}}{2}\right)^2 + (\tau_{zx})^2} \qquad (12)$$

$$2\sigma^2 = (\sigma_a - \sigma_b)^2 + (\sigma_b - \sigma_c)^2 + (\sigma_c - \sigma_a)^2 \qquad (13)$$

Equation (7), (8), (9) represent the relation for normal stresses $\sigma_{xx}$, $\sigma_{yy}$, $\sigma_{zz}$ with the normal strains $\varepsilon_{xx}, \varepsilon_{xx}, \varepsilon_{xx}$ in x, y, z direction respectively. E is young's modulus, $\alpha$ is the coefficient of thermal expansion and $\nu$ is the Poisson's ratio of the biological media.

Assuming the tissue and tumor material to be ductile, we consider Von-Mises criteria of failure for the study. According to Von-Mises criteria, material fails when the shear strain energy per unit volume reaches the maximum strain energy of the body. $\sigma$ is the Von-Mises stress developed in the body due to $\sigma_a, \sigma_b, \sigma_c$ which are the principal stresses in three mutually perpendicular directions. Equation (10), (11), (12) represent the relation for $\sigma_a, \sigma_b, \sigma_c$ with $\tau_{xy}\tau_{yz}\tau_{zx}$ which are the shear stresses on the respective planes, Equation (13) shows the relation for Von-Mises stress with principal stresses.

## 2.3 Numerical scheme and validation

Finite element based COMSOL Multiphysics software was used to predict the temperature stress for 180 mins with a time step of 0.1min at a tolerance of 0.01. Grid independence test was done for the present model by determining the temperature variation at a point which is 10 mm away from the tumor centre in positive x-direction for 40 mins while performing the simulation of cryosurgery using a probe of 6.9mm diameter. It has been found that the temperature became constant from grid count of 146215. All the results were extracted for this mesh size.

To validate the present simulation results, 3-D numerical study was done for the experiment conducted by Seifart et al. [8]. Here, the authors conducted the cryosurgery experiment on a pig in-vivo. Authors observed the variation of temperature with respect to time at different locations in the tissue around the cryoprobe for different time spans while in different working conditions. Referring to this experiment, temperature at a distance of 10mm from the centre of the tumor with an 8mm diametrical probe was determined for a time span of 40 mins as shown in the fig. 3. A satisfactory result was obtained for the simulation which showed an average error of $\pm 3^0 C$.

<Fig. 3>

3. Results and Discussion

Following results were obtained by numerically analysing the thermal and mechanical behaviour of the cryosurgery process on the mentioned domain model. In the following thermal contours, $-40^0 C$ isotherm has been represented to identify the confirmed cell death region and there is a $10^5 \ N/m^2$ stress curve in the stress contours to identify the pain tolerable region in the tissue.

Fig. 4(a), 4(b) represent the temperature distribution with respect to space in x and y direction lines passing through the tumor centre at five different time steps starting from 0 mins, 20 mins, 40 mins, 120 mins, 180 mins. Similarly, fig. 4(c), 4(d) represent the stress distribution in x and y direction respectively at the previously mentioned time steps. It can be observed that the temperature is asymptotically increasing as the distance increases from the tumor surface. It is also seen that the stress decreases with increase in the space coordinates but suddenly increases at the tumor surface due to sudden change in the mechanical properties of the domain at the tumor and healthy tissue interface. From fig. 4(a), 4(b) it can also be seen

that the rate of temperature-drop is decreasing as the distance increases from the tumor centre.

<Fig. 4>

<Fig. 5>

Fig. 5(a), 5(b) represents the variation of temperature and stress with respect to time at three different locations in the domain, located at 5 mm, 10 mm and 15 mm distance away from the tumor centre in x direction. First specified location is inside the tumor, second on the tumor-healthy tissue interface and the third in the healthy tissue. It can be seen that temperature decreased and stress increased with time at all the specified locations. Fig. 5(a) shows that initially the heat transfer in the domain is rapid and then reaches steady state at around 50 mins at all the specified locations. Similarly, from fig. 5(b), it can be noted that even stress generation was rapidly increasing initially in the tumor (at the location 5mm away from the tumor centre) and then reached the steady state at around 50 mins. But, the healthy tissue (at the location 15mm away from the tumor centre) has experienced a rapid increase in the stress at around 50 mins and continued till the end of the process. This is due to high young's modulus (E) of frozen healthy tissue. Fig. 5(c) when compared with fig. 5(a), it shows that the strain of the tissue is proportional to the temperature at that particular time and location. Fig. 5(d) represents the variation of stress with respect to strain at all the three locations, it can be seen that the tumor experienced more stress and strain when compared to the surrounding healthy tissue.

<Fig. 6>

Fig. 6 represents the top view and front view of the temperature contour (left) and stress contour (right) of xy, xz planes which are passing through the tumor centre at the time steps of 0min, 20min, 40min, 80min, 120 min and 180 min. Black curved line in fig. 6 represents $-40^0 C$ isotherm and $10^5 \ N/m^2$ stress curve in temperature and stress contours respectively. Referring to these temperature and stress contours, it can be noted that constant temperature and stress lines are growing rapidly from 0 to 80 mins but there is no remarkable change further in the tumor tissue. Fig. 7 represents the deformation of the tumor tissue at the mentioned time steps. Here, tumor tissue is going through a rapid deformation till 40 mins. From both fig. 6 and fig. 7 it is observed that the region around the probe has experienced very high stress. Fig. 7 also shows that the tumor has undergone higher deformation than the healthy tissue.

<Fig. 7>

### 3.1 Parametric study

A parametric study has been done by varying the tumor size, probe size and the tumor location individually and keeping the other parameters unchanged. This study helps in recognizing and understanding the effect of each and every parameter on the cryosurgery process.

### 3.1.1 Effect of tumor size

Here, study was done by changing the tumor size -20 mm, 22 mm, 25 mm, 29 mm, for a fixed probe diameter of 6.9 mm.

<Fig. 8>

Fig. 8 shows the heat flux on the right face of the tumor at 40 mins for the mentioned tumor sizes. Heat flux suddenly increased initially as there was a rapid drop in the temperature in the beginning and then remained constant from around 50 mins. It is also observed that, as the tumor size increases, heat flux through the tumor face decreases as the tumor size increases as the volume of heat absorption increases which results in a decrease in the heat flux.

<Fig.9>

Fig. 9(a) shows the variation of temperature with respect to time and fig. 9(b) shows the variation of stress with respect to time at a point 10mm away from the tumor centre in x-direction for the mentioned sizes of tumor. It is observed that the rate of heat transfer increased as the tumor size decreased and then reached a steady state and therefore there was high thermal stress in the beginning and then reached a constant value as the heat transfer reached the steady state.

Fig. 10 represents top view and front view of temperature (top) and stress distribution (bottom) contour of xy, xz planes which are passing through the tumor centre in the increasing order of mentioned tumor sizes from left to right at the time step of 40mins. Even these contours demonstrate that when tumour size increases, heat transmission and stress distribution rapidly diminish. Black curved line in the figure represents the same as mentioned earlier.

<Fig.10>

### 3.1.2 Effect of variation of probe diameter

Behaviour of the cryosurgery process was examined by varying the probe diameter 5mm, 6mm, 6.9mm, 8mm for constant tumor size of 20mm.

Here, fig.11 (a) and fig. 11(b) represent the change in temperature and stress with respect to time for the mentioned probe sizes. It is observed that the rate of heat transfer and stress induced in the domain increased with increase in the diameter of the probe. It can be seen that the rate of temperature drop is high till around 50 mins and gradually decreases as the time passes. From the fig. 11(b) it can be seen that as the time passed, initially stress increased rapidly to a maximum value and then decreased by some extent and then further passed till the end of the process with no remarkable change.

<Fig. 11>

<Fig. 12>

In fig. 12, top view and front view of temperature (top) and stress (bottom) distribution of xy, xz planes passing through the tumor centre at 40 mins for the given probe sizes have been extracted. From the temperature and stress contours it can be noted that heat transfer and stress generation have increased with increase in the diameter of the probe. Black curved line in the figure represents the same as mentioned earlier.

### 3.1.3 Effect of change of tumor location

Process was also studied by placing the 20 mm tumor at different locations in the tissue when treated with a 6.9mm diametrical probe. Following are the four different kinds of tumors based on their location.

Internal tumor - Tumor is 5mm below the top surface of the tissue.

Surface tumor - Tumor's top surface coincides with the top surface of tissue.

Partial tumor - Half the tumor is outside the tissue.

External tumor - Tumor is completely outside the tissue.

Fig. 13(a) represents the variation of temperature with respect to time and fig. 13(b) represents stress variation with respect to time at a point 10mm away from the tumor centre in x direction when the tumor is placed at the specified locations in the tissue. It is observed that, rate of temperature drop increased rapidly as the tumor moved out of the tissue. Even here it is observed that both heat transfer and stress reached steady state at around 50 mins. When fig. 13(b) is observed, it can be noted that, as the tumor moved outward, maximum stress experienced by the tumor increased, but the stress gradually dropped as the time passed. By the end of the process, it can be seen that the surface tumor experienced the highest stress and the external tumor experienced the lowest stress among all kinds of tumors.

<Fig. 13>

4. Conclusions

Few important inferences can be drawn out by observing the results. The study reveals that there is a significant drop in temperature in the tissue alone till 50 mins of treatment after which it reaches steady state. Results also show that the temperature and stress are varying dominantly till around 50 mins in the tumor with no remarkable change further. It should be noted that following 40 mins of the process, stress is developing rapidly in the healthy tissue and therefore the process should be strictly limited within this time limit for preventing severe pain in the tissue. By the end of the process, observations predict a little amount of deformation in the healthy tissue region and moderate amount of deformation in the tumor region. The tumor experiences a very high amount of pain when compared to the surrounding tissue as it can be observed that the rate of decrease in temperature blows up in the tumor region and rapidly drops down in the surrounding healthy tissue region. Results also indicate that surrounding healthy tissue might experience severe pain by the end of the process. Parametric study shows that, for a given probe diameter, heat transfer rate decreases rapidly as the tumor size increases. It can also be seen that, for a given tumor size, as the probe size increases heat transfer rate increases. Analysis also shows that as the tumor size increases, induced stress decreases heavily but it is to be noted that, as the probe size increases, there is no recognizable change in stress distribution in the tumor tissue. Parametric study also reveals that, as the tumor moves out of the tissue, it can be seen that both heat transfer and maximum stress are increasing rapidly. This can infer that the cryosurgery process becomes simpler as the tumor moves outward of the tissue but at the same time, pain increases as the tumor moves outward. These results and inferences might extend the understanding of researchers and surgeons and give an idea about the duration of surgery, pain experienced by the patient and the effect on the surrounding healthy tissue during and after the cryosurgery.

**Nomenclature**

**Notations**                     **Descriptions**

T                                 Temperature, $degC$

$K$                               Thermal conductivity of tissue

Q                                 Heat generation

T                                 Time

C                                 Specific heat

$\theta$                          Mass fraction

m                                 Mass fraction while phase change

$\sigma$                          Stress

$\varepsilon$                     Strain

$\propto$                         Coefficient of thermal expansion

$\nu$                              Poisson's ratio

E                                 Young's modulus

$l$                               length

**Subscripts**

xx                                x direction

yy                                y direction

zz                                z direction

1                                 liquid phase

2                                 solid phase

b                                 blood perfusion

m                                 metabolism

a                                 1st principal stress direction

b                                                  $2^{nd}$ principal stress direction

c                                                  $3^{rd}$ principal stress direction

# References

1. WHO. *Cancer*. 2020. https://www.who.int/news-room/fact-sheets/detail/cancer

2. Sharma, D. L. *ICMR-NCDIR National Cancer Registry Programme estimates 12% increase in cancer cases in the country by 2025*. 2020 https://ncdirindia.org/All_Reports/Report_2020/PB/Press_release.pdf

3. NCI. *What Is Cancer?*. https://www.cancer.gov/about-cancer/understanding/what-is-cancer

4. Siegel, R., Naishadham, D., and Jemal, A. Cancer statistics. 2012. *CA Cancer J Clin*. 2012; 62(1): 10–29.

5. Mazur, P. Cryobiology: The freezing of biological systems. *Science.*1970; 168(3934): 939–949.

6. Chua, K.J., Zhao, X., and Chou, S. K. Effects of crucial parameters on the freezing delivery in the cryosurgical system. *Appl Therm Eng*. 2013; 51(1–2): 734–741.

7. Rubinsky, B. Cryosurgery. *Annurev.bioeng*. 2000; 2: 157–187.

8. Seifert, J. K., Gerharz, C. D., Mattes, F., Nassir, F., Fachinger, K., Beil, C., and Junginger, T. A pig model of hepatic cryotherapy. In vivo temperature distribution during freezing and histopathological changes. *Cryobiology*. 2003; 47(3): 214–226.

9. Zhmakin, A. I. Fundamentals of cryobiology: physical phenomena and mathematical models. 1st Edition. Springer Berlin, Heidelber; 2009.

10. Khademi, R., Mohebbi-Kalhori, D., and Razminia, A. Thermal analysis of a tumorous vascular tissue during pulsed-cryosurgery and nano-hyperthermia therapy: Finite element approach. *Int J Heat Mass Transf*. 2019; 137: 1001–1013.

11. Chua, K.J. Computer simulations on multiprobe freezing of irregularly shaped tumors. *Comput Biol Med*. 2011; 41(7): 493–505.

12. Stanke, K. M., and Ivanec, D. Pain threshold - Measure of pain sensitivity or social behavior?. *Psihologija*. 2016; 49(1): 37–50.

13. Silver, F. H., Christiansen, D. L. Biomaterials Science and Biocompatibility. 1st Edition. Springer New York, NY; 1999.

14. Fischer, A. A. Pressure algometry over normal muscles. Standard values, validity and reproducibility of pressure threshold. *Pain*. 1987; 30(1): 115-126.

15. Park, G., Kim, C. W., Park, S. B., Kim, M. J., and Jang, S. H. Reliability and Usefulness of the Pressure Pain Threshold Measurement in Patients with Myofascial Pain. *Ann Rehabil Med*. 2011; 35(3): 412-417.

16. Gage, A. A., and Baust, J. G. Cryosurgery for Tumors-A Clinical Overview. *Technol Cancer Res Treat*. 2004; 3(2): 187–199.

17. Pennes, H. H. Analysis of tissue and arterial blood temperatures in the resting human forearm. *J. Appl. Phycol*. 1948; 1(2): 93-122.

18. Wissler, E. H. 50 years of JAP Pennes' 1948 paper revisited. *J. Appl. Phycol*. 1998; 85(1): 35-41.

19. Lipkin, M., and Hardy, J. D. Measurement of Some Thermal Properties of Human Tissues. *J. Appl. Phycol.* 1954; 7(2): 212-217.

20. He, X., and Bischof, J. C. Analysis of thermal stress in cryosurgery of kidneys. *J Biomech Eng*. 2005; 127(4): 656–661.

21. Rabin, Y., and Steif, P. S. Analysis of Thermal Stresses around a Cryosurgical Probe. *Cryobiology*. 1996; 33(2): 276-290.

22. Rabin, Y. and Steif, P. S. Thermal stress modelling in cryosurgery. *Int J Solids Struct.* 2000; 37(17): 2363-2375.

23. Shitzer, A. Cryosurgery: Analysis and experimentation of cryoprobes in phase changing media. *J Heat Transfer*. 2011; 133(1): 011005-16.

24. Zhao, X., and Chua, K. J. Studying the thermal effects of a clinically-extracted vascular tissue during cryo-freezing. *J Therm Biol*. 2012; 37(8): 556–563.

25. Chua, K. J. Fundamental experiments and numerical investigation of cryo-freezing incorporating vascular network with enhanced nano-freezing. *Int. J. Therm. Sci*. 2013; 70: 17–31.

26. Li, E., Liu, G. R., Tan, V., and He, Z. C. An efficient algorithm for phase change problem in tumor treatment using αFEM. *Int. J. Therm. Sci.* 2010; 49(10): 1954–1967.

27. Zhao, X., and Chua, K. J. Investigating the cryoablative efficacy of a hybrid cryoprobe operating under freeze-thaw cycles. *Cryobiology*. 2013; 66(3): 239–249.

28. Liu, J., Zhou, Y., Yu, T., Gui, L., Deng, Z., and Lv, Y. Minimally invasive probe system capable of performing both cryosurgery and hyperthermia treatment on target tumor in deep tissues. *Minimally Invasive Therapy and Allied Technologies*. 2004; 13(1): 47–57.

29. Paul, A., and Paul, A. Thermomechanical analysis of a triple layered skin structure in presence of nanoparticle embedding multi-level blood vessels. *Int J Heat Mass Transf*. 2020; 148: 119076.

30. Paul, A., and Paul, A. In Vitro Thermal Assessment of Vascularized Tissue Phantom in Presence of Gold Nanorods during Photo-Thermal Therapy. *J Heat Transfer*. 2020; 142(10): 101201-14.

31. Paul, A., and Paul, A. Thermo-mechanical assessment of nanoparticle mixed vascular tissues under pulsed ultrasound and laser heating. *Int. J. Therm. Sci.* 2021; 163: 106815.

32. Chua, K. J., Chou, S. K., and Ho, J. C. An analytical study on the thermal effects of cryosurgery on selective cell destruction. *J Biomech*. 2007; 40(1): 100–116.

33. Kumar, A., Kumar, S., Katiyar, V. K., and Telles, S. Phase change heat transfer during cryosurgery of lung cancer using hyperbolic heat conduction model. *Comput Biol Med*. 2017; 84: 20–29.

34. Nakayama, A., Kuwahara, Y., Iwata, K., and Kawamura, M. The limiting radius for freezing a tumor during percutaneous cryoablation. *J Heat Transfer*. 2008; 130(11): 1–6.

35. Ramajayam, K. K., and Kumar, A. A novel approach to improve the efficacy of tumour ablation during cryosurgery. *Cryobiology*. 2013; 67(2): 201–213.

36. Shi, J., Chen, Z., and Shi, M. Simulation of heat transfer of biological tissue during cryosurgery based on vascular trees. *Appl Therm Eng*. 2009; 29(8–9): 1792–1798.

37. Zhao, G., Zhang, H. F., Guo, X. J., Luo, D. W., and Gao, D. Y. Effect of blood flow and metabolism on multidimensional heat transfer during cryosurgery. *Med Eng Phys*. 2007; 29(2): 205–215.

38. Li, X., Luo, P., Qin, Q. H., and Tian, X. The phase change thermoelastic analysis of biological tissue with variable thermal properties during cryosurgery. *Journal of Thermal Stresses*. 2020; 43(8): 998–1016.

39. COMSOL Multiphysics Reference Manual; 1998

## List of Table

Table 1: Thermal and mechanical properties of tissue and tumor [38]

|  | Tumor | | Healthy tissue | | Blood |
|---|---|---|---|---|---|
|  | **Frozen** | **Unfrozen** | **Frozen** | **Unfrozen** | - |
| E(Pa) | $7.63 \times 10^9$ | $7.63 \times 10^6$ | $2.3 \times 10^9$ | $2.29 \times 10^7$ | - |
| $\nu$ | 0.33 | 0.48 | 0.33 | 0.48 | - |
| $\rho(kg/m^3)$ | 1660 | 1660 | 1000 | 1000 | 1060 |
| K(W/mk) | 2.25 | 0.55 | 0.38 | 0.45 | - |
| C(J/kgK) | 2540 | 4200 | 1800 | 3600 | 3770 |
| $Q(W/m^3)$ | 0 | 4200 | 0 | 380 | - |
| w(1/s) | 0 | 0.02 | 0 | 0.0005 | - |
| $\alpha(1/K)$ | $6 \times 10^{-5}$ | $7.3 \times 10^{-5}$ | $6 \times 10^{-5}$ | $7.32 \times 10^{-5}$ | - |

# List of Figures

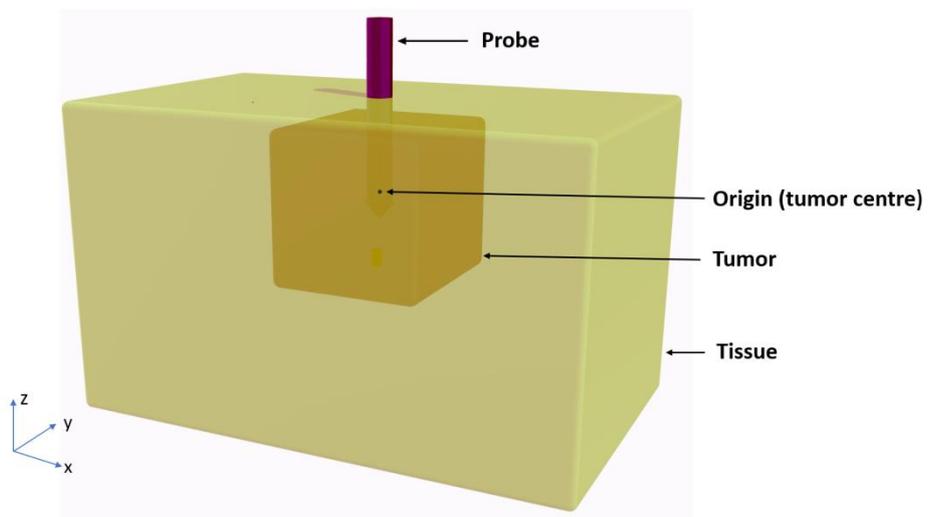

Figure 1: Physical model representing healthy tissue (cuboid), tumor (internal cube) and the cryoprobe (cylinder with conical tip) where origin is at the tumor centre.

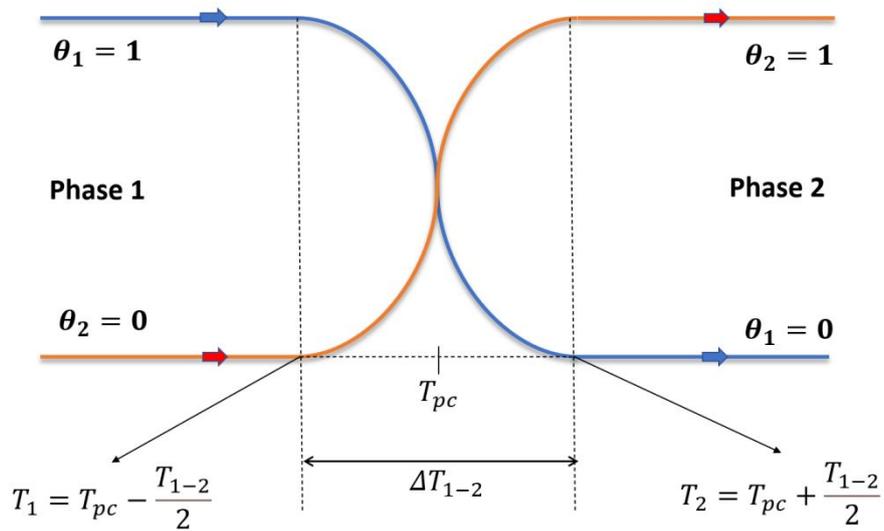

Figure 2: Phase transition model, blue line represents the reduction of the liquid phase and the red line shows the increase in the solid phase where $\theta_1, \theta_2$ are the mass fractions of liquid and solid phase of biological media and $T_{pc}$ is the phase change temperature. Region between the dotted lines represent the exact phase transition process from $T_1$ to $T_2$.

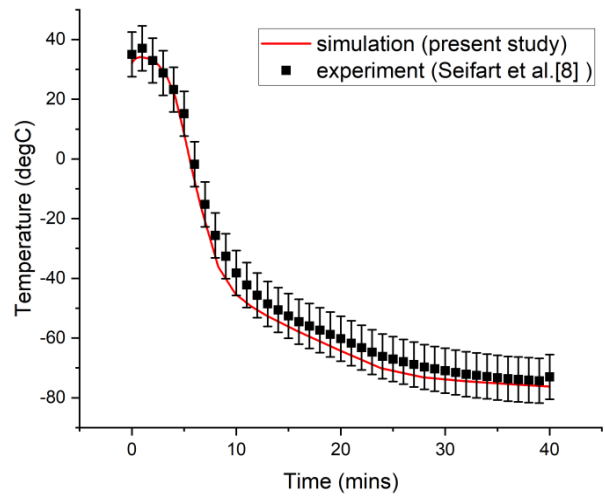

Figure 3: Comparison of predicted and experimental thermal history at a location 10mm away from the tumor center along x axis during the cryosurgery.

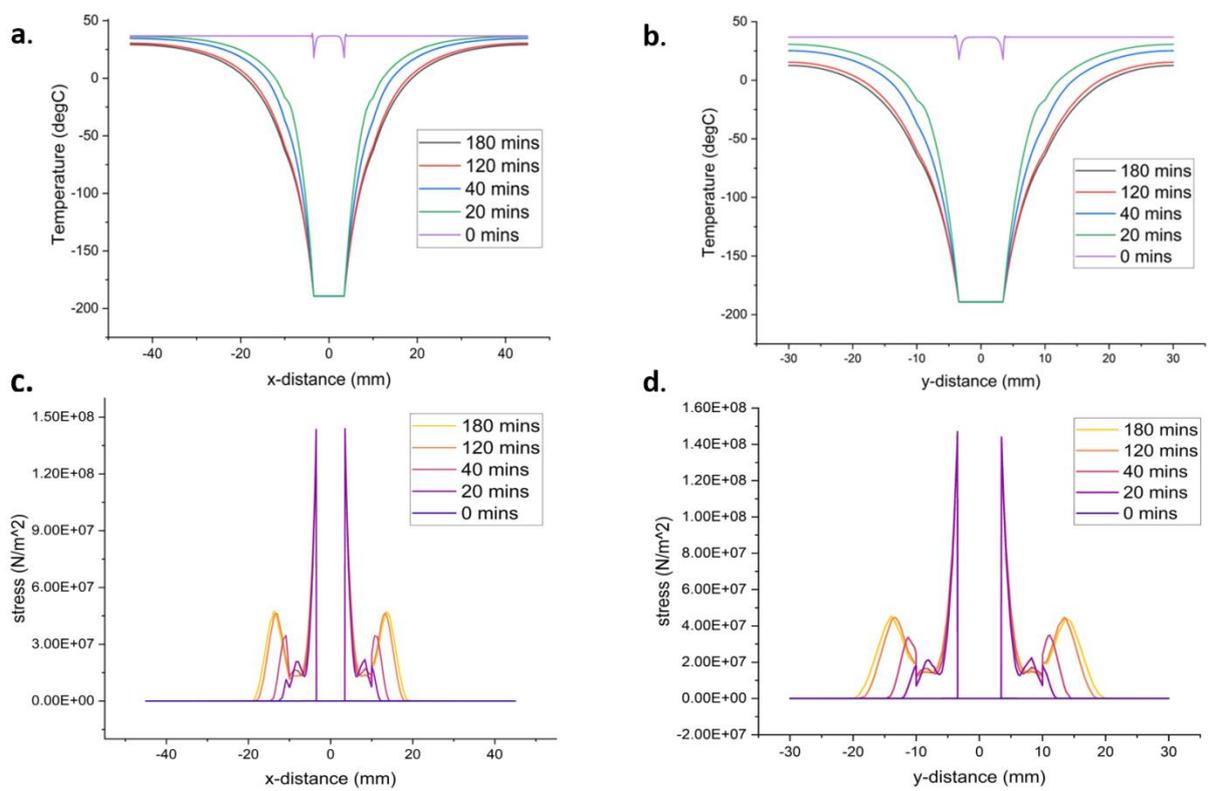

Figure 4: Predicted (a) temperature variation along x axis (y=0, z=0), (b) temperature variation along y axis (x=0, z=0), (c) stress variation along x axis (y=0, z=0), (d) stress variation along y axis (x=0, z=0), at different times of cryoablation.

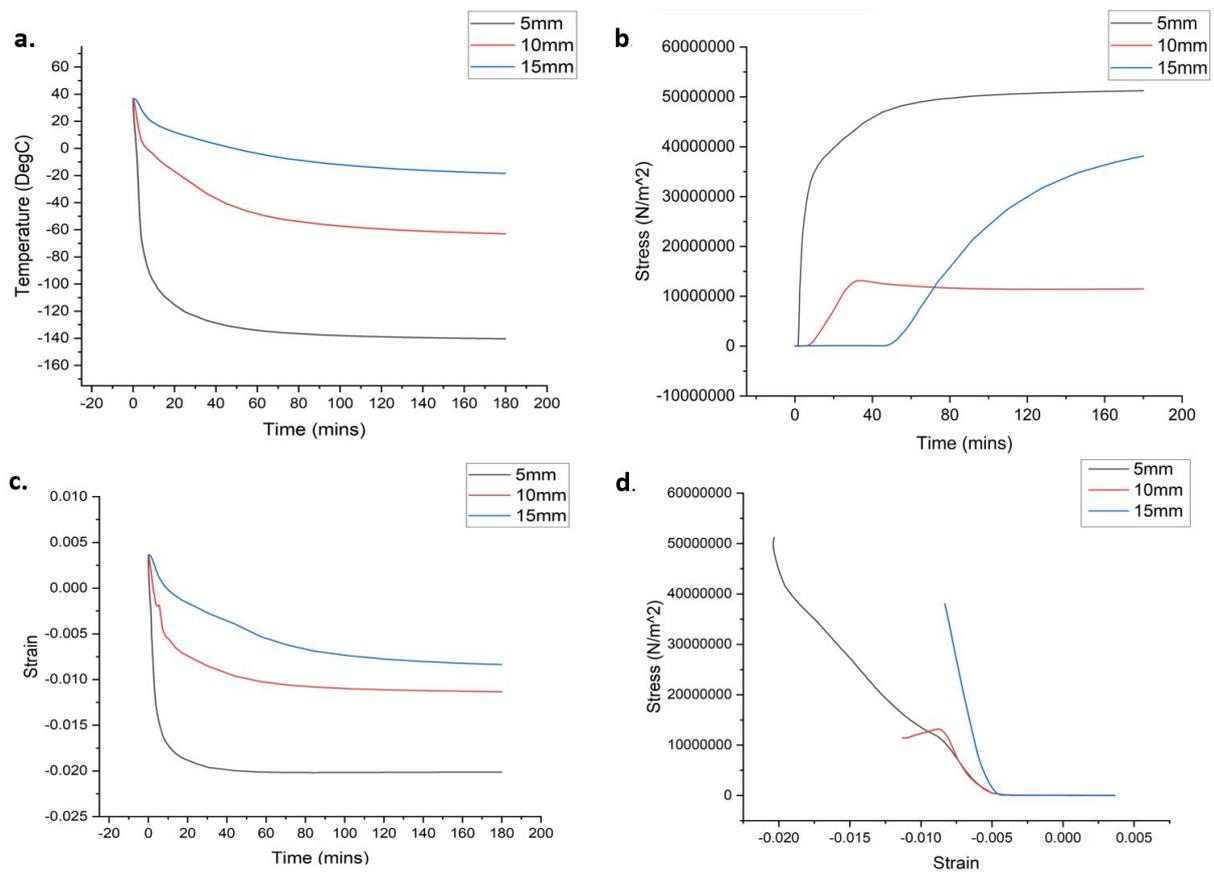

Figure 5: (a) Temperature variation with time, (b) Stress variation with time, (c) Strain variation with time, (d) Stress variation with respect to strain, at different locations which are at the mentioned distances on positive x-axis (y=0, z=0) during cryosurgery.

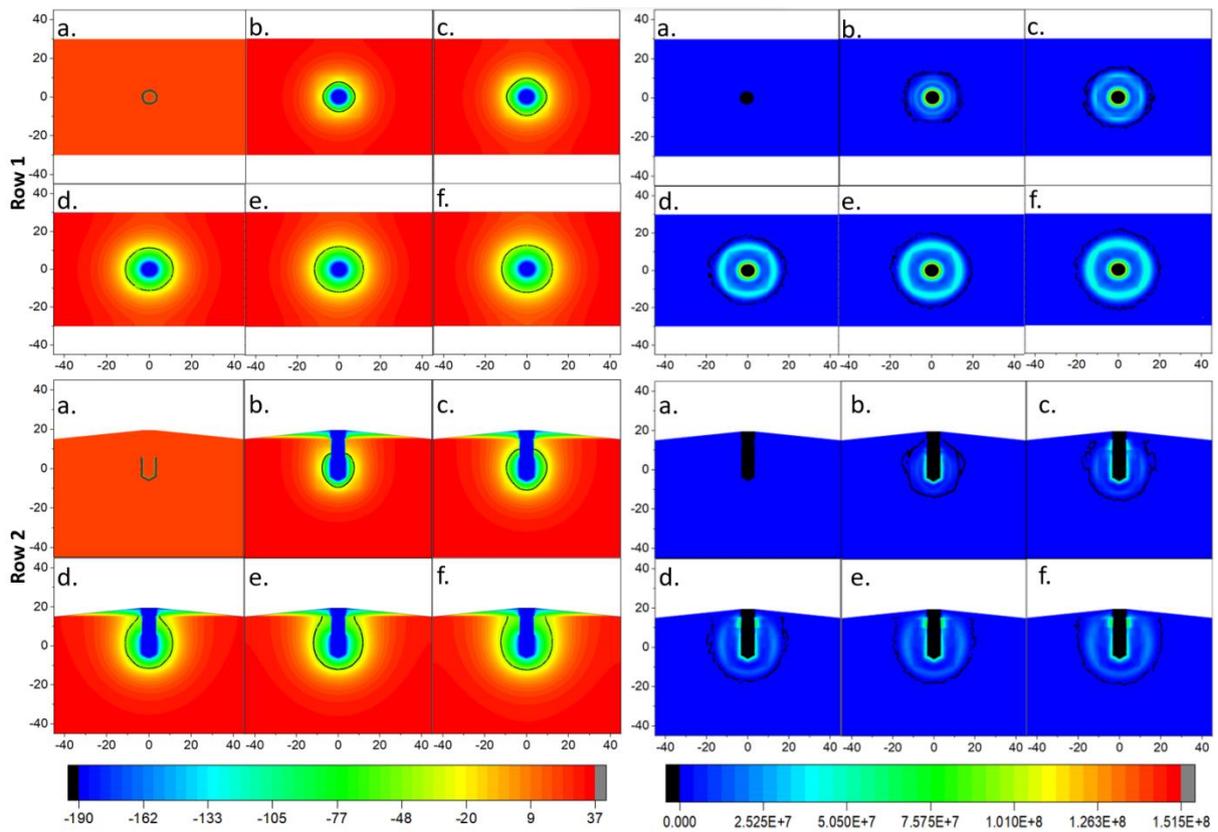

Figure 6: Temperature ($degC$) (left column) and stress ($N/m^2$) (right column) distribution on xy plane at z=0 (Row 1) and xz plane at y=0 (Row 2) at (a) 0 mins, (b) 20 mins, (c) 40 mins, (d) 80 mins, (e) 120 mins, (f) 180 mins.

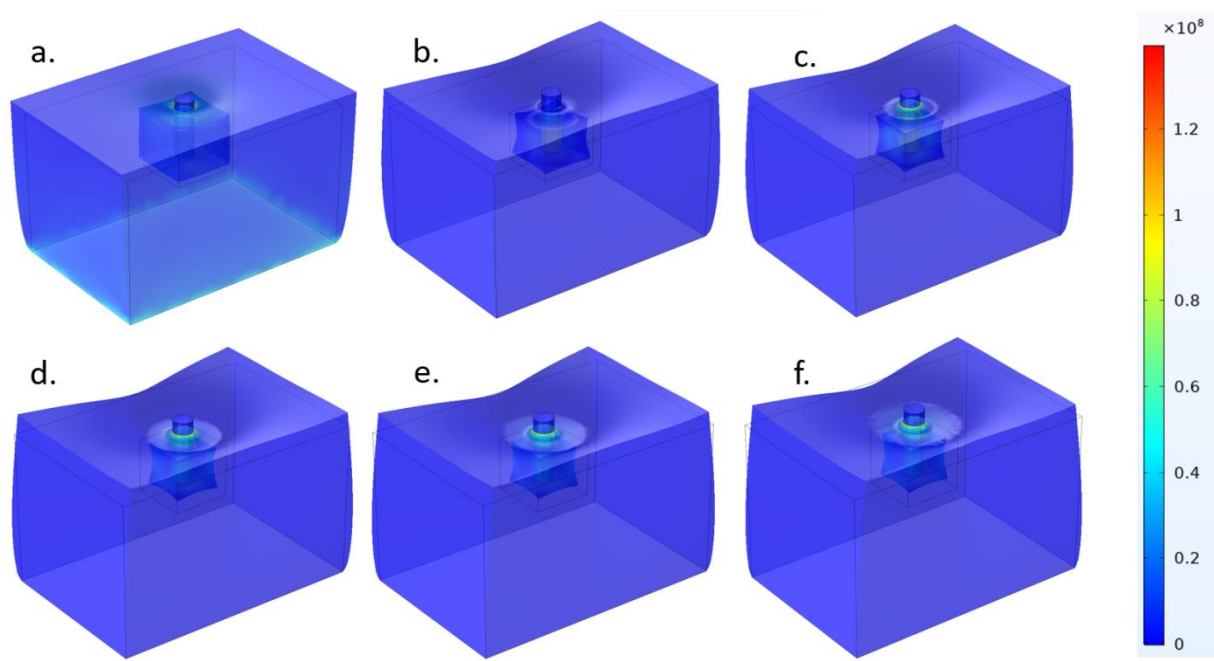

Figure 7: Tumor-tissue deformation and induced stress ($N/m^2$) in the tissue at different time of cryoablation (a) 0 mins, (b) 20 mins, (c) 40 mins, (d) 80 mins, (e) 120 mins, (f) 180 mins.

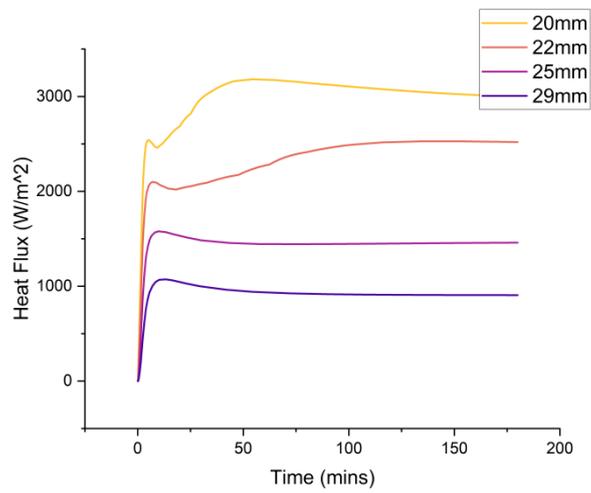

Figure 8: Variation of surface heat flux with respect to time on the right face of tumor when viewed in positive y-direction for different sizes of tumor.

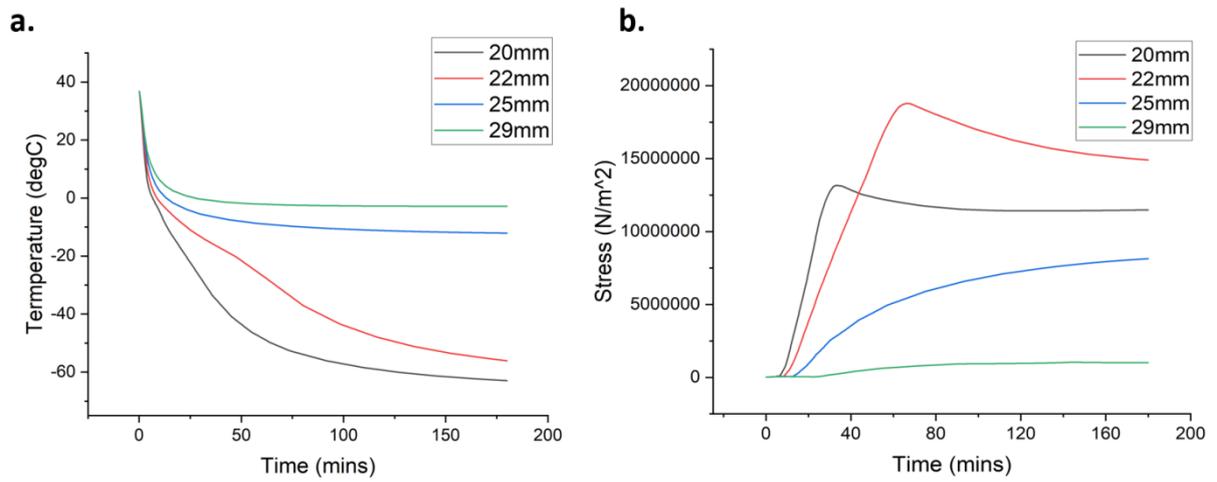

Figure 9: (a) Variation of temperature with time and (b) Variation of stress with time, at the location (10,0,0) for different sizes of tumor.

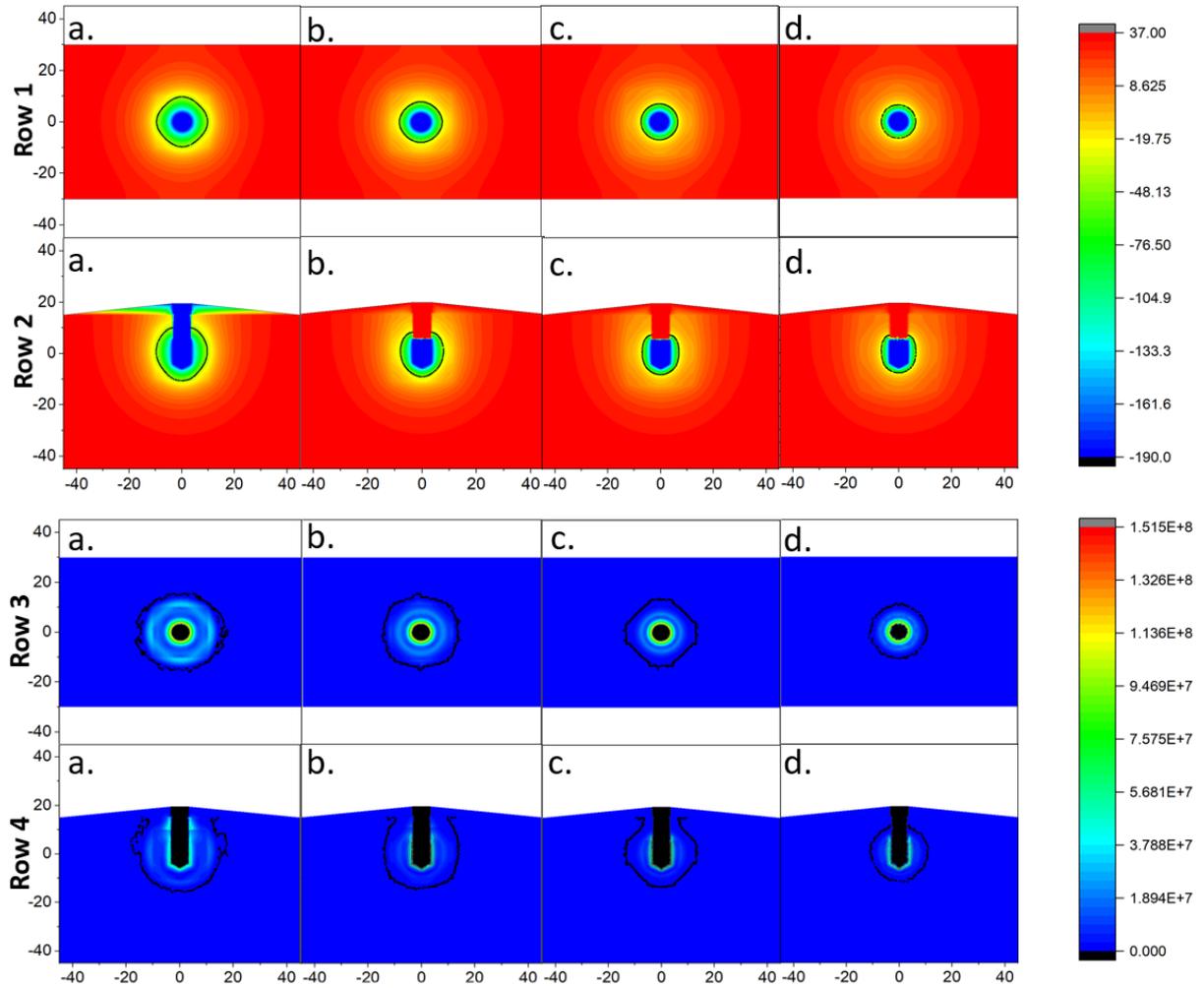

Figure 10: Row 1, row 2 represent the temperature ($degC$) distribution on xy plane (z=0), xz plane (y=0) respectively and row 3, row 4 represent the stress ($N/m^2$) distribution on xy plane (z=0), xz plane (y=0) respectively at 40mins of cryosurgery for the tumor sizes (a) 20mm, (b) 22mm, (c) 25mm, (d) 29mm.

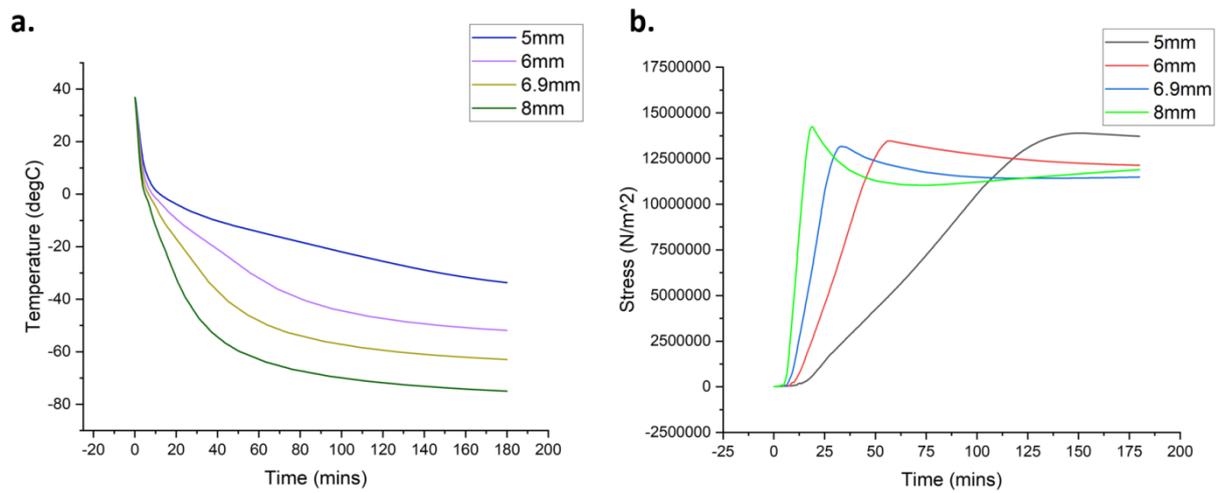

Figure 11: (a) Variation of temperature with time at the location (10mm,0,0) for different probe sizes as shown. (b) variation of stress with time at the location (10mm,0,0) for different probe sizes.

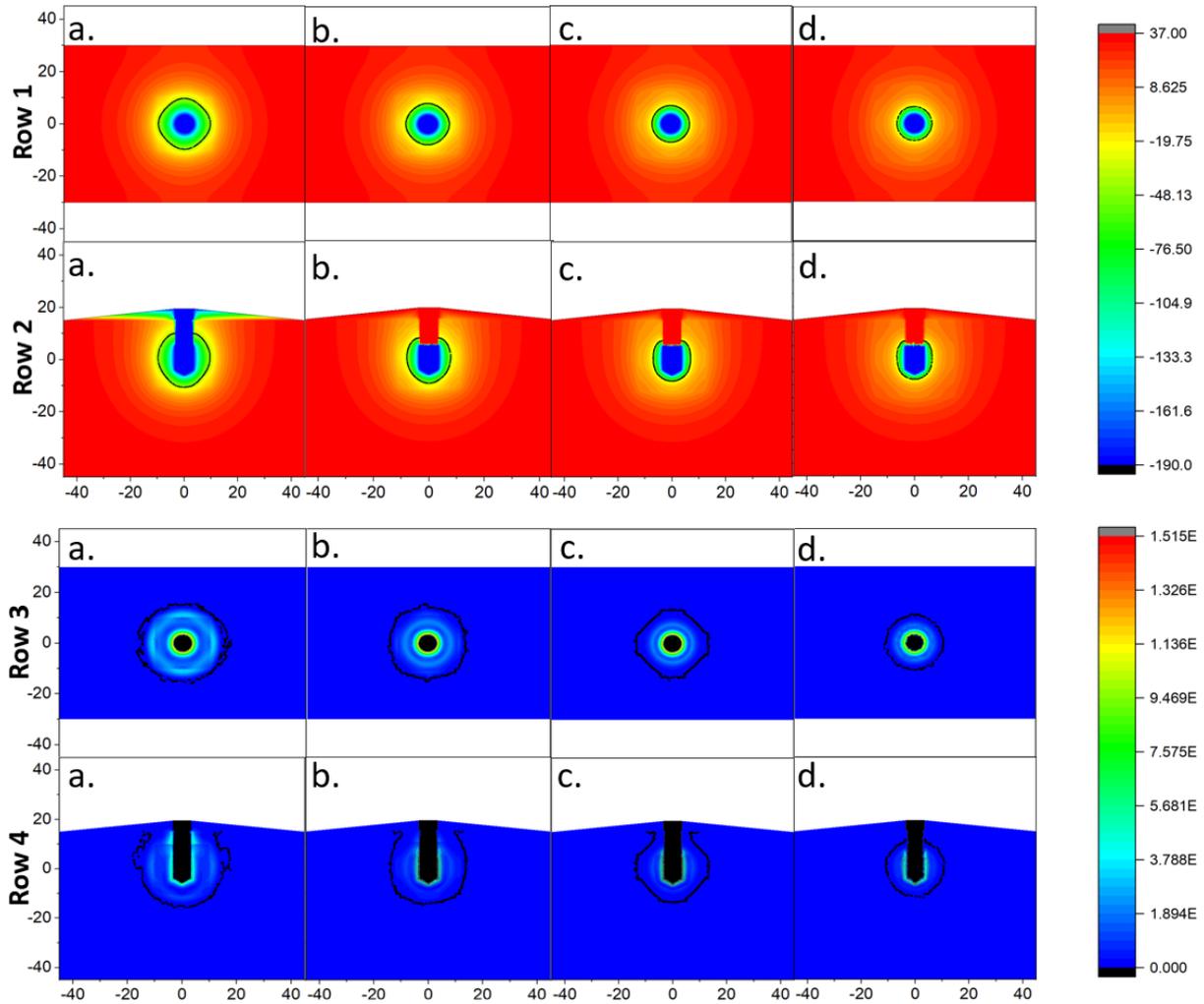

Figure 12: Temperature (*degC*) distribution on xy plane at z=0 (row 1) and xz plane at y=0 (row 2) and stress ($N/m^2$) distribution on xy plane at z=0 (row 3) and xz plane at y=0 (row 4), at 40mins of cryosurgery for the probe size- (a) 5mm, (b) 6mm, (c) 6.9mm, (d) 8mm.

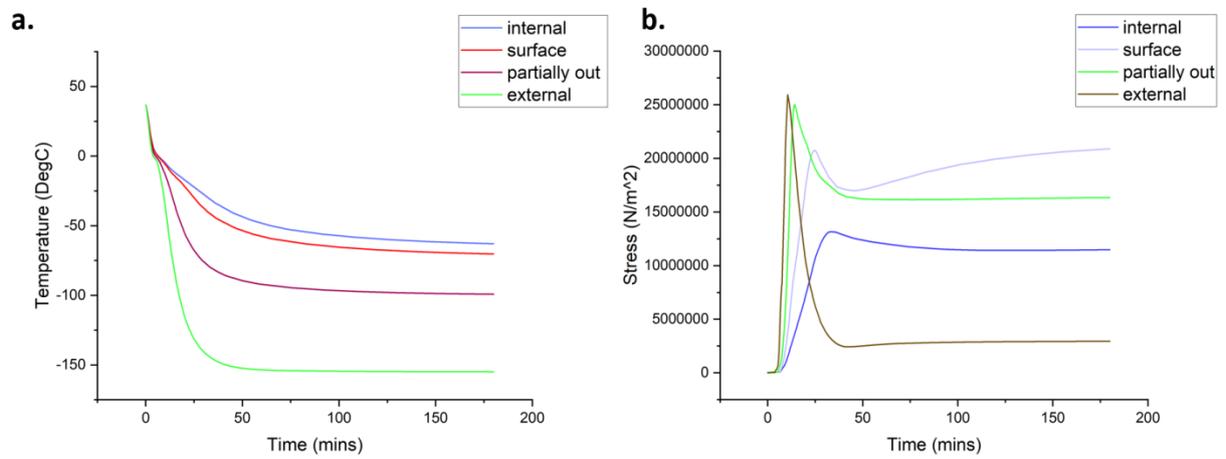

Figure 13: (a) Temperature variation with time at the location (10mm,0,0) (b) Stress variation with time at the location (10mm,0,0) for different positions of tumor.